\documentclass[final,3p,times]{elsarticle}

\usepackage{amsfonts}
\usepackage{amsbsy}
\usepackage{amssymb}
\usepackage{amsmath}
\usepackage[usenames]{color}
\usepackage{mathptmx}
\usepackage{graphicx}
\usepackage{mathtools}
\usepackage{natbib}
\usepackage{hyperref}
\usepackage{amsthm}

\newtheorem{proposition}{Proposition}

\journal{Applied Mathematics Letters}

\begin{document}
	
\begin{frontmatter}
\title{Exact Solutions to a Family of Complex Ginzburg--Landau Equations with Cubic--Quintic Nonlinearity}
		
\author{Vassil M. Vassilev}
\ead{vasilvas@imbm.bas.bg}
\address{Institute of Mechanics, Bulgarian Academy of Sciences, Acad. G. Bonchev St., Block 4, 1113 Sofia, Bulgaria}

\begin{abstract}
In these notes, using the method of differential constraints, novel exact kink-like solutions are obtained for certain classes of complex Ginzburg--Landau equations with cubic-quintic nonlinearity. The foregoing solutions are presented in terms of the Lambert $W$ function.
\end{abstract}
		
		
\end{frontmatter}		
	
\section{Introduction}
\label{Sec:1}
A wide variety of processes and phenomena studied in the theory of phase transitions, laser and plasma physics, nonlinear optics, physics of electrical transmission lines and many other branches of natural sciences are modelled on the ground of the so-called \textit{complex Ginzburg--Landau equations} (see, e.g., \cite{Agrawal2019,Akhmediev1997,Aranson2002,Biswas2007,Hohenberg2015,Kengne2022,Liu2019,Malomed2021,vanSaarloos1992,vanSaarloos2003}).
Equations of this type are frequently called also \textit{derivative nonlinear Schr\"{o}dinger equations} since they maid be thought of as generalizations of the \textit{standard nonlinear Schr\"{o}dinger equation} extended by appending different kinds of linear and nonlinear terms in order to capture the interplay between dispersive and nonlinear effects (see, e.g., \cite{Kengne2022,Liu2019,Cross1993} and the references therein). 

In the present work, we consider the following family 
\begin{equation} \label{CCQGLE}
	i\frac{\partial U}{\partial x}-\tau _{R}U\frac{\partial \left\vert
		U\right\vert ^{2}}{\partial t}+\left( \frac{\alpha }{2}-i\beta \right) \frac{%
		\partial ^{2}U}{\partial t^{2}}-i\delta U+\left( \gamma -i\varepsilon
	\right) \left\vert U\right\vert ^{2}U+\left( \lambda -i\mu \right) \left\vert
	U\right\vert ^{4}U=0
\end{equation}
of complex Ginzburg--Landau equations with cubic-quintic nonlinearity. Here, $U(x, t)$ is a complex-valued function of the real variables $x$ and $t$, while $\tau _{R} \neq 0$, $\alpha \neq 0$, $\beta \neq 0$, $\gamma$, $\delta$, $ \varepsilon $, $\mu$ and $\lambda$ are real constants. The physical meaning of these parameters, the sought function $U(x, t)$ and the independent variables $x$ and $t$ will not be commented on here because it depends on the theory within the framework of which equation (\ref{CCQGLE}) is regarded.

Equations of the form (\ref{CCQGLE}) have been studied in a series of works, especially in the context of the nonlinear optics (see, e.g., \cite{Malomed2021,Agrawal1992,Akram2018,Akhmediev1995,Akhmediev1996,Izgi2023,Kudryashov2022a,Nisha2020,Petviashvili1984, Uzunov2014,Uzunov2018,Uzunov2020,Wazwaz2006,Yan2019}).
It is worth noting that, as is pointed out by Malomed \cite{Malomed2021}, the cubic-quintic nonlinearity was first introduced by Petviashvili and Sergeev \cite{Petviashvili1984} in 1984.
It has been established that under certain restrictions on the values of the parameters, equations of the form (\ref{CCQGLE}) admit exact solutions of different types, a wide variety of integration techniques being used for that purpose.

The aim of the current work is to reveal new exact solutions to equations belonging to the foregoing family using the method of differential constraints (see, e.g., \cite[Chapter 4]{Andreev1998}, \cite{Olver1994} and the references therein).
This study may be regarded as a continuation of the work by Nisha \textit{et al.} \cite{Nisha2020} in which the authors have presented a new type of kink-like solutions to equations of form (\ref{CCQGLE}) named Lambert $W$-kinks as they are expressed in terms of the Lambert $W$ function \cite{Corless1996}.
In Section \ref{Sec:4} of the present notes, four new classes of equations of the form (\ref{CCQGLE}) are identified that admit Lambert $W$-kink solutions. 

\section{Reduction to Ordinary Differential Equations}
\label{Sec:2}

Equation (\ref{CCQGLE}) admits reduction to a system of ordinary differential equations through the ansatz
\begin{equation} \label{ansatz}
	U\left( x,t\right) =\rho \left( \xi \right) \exp \left[ i\left( \phi \left( \xi
	\right) -kx\right) \right], 
\end{equation}
where $\rho(\xi)$ and $\varphi(\xi)$ are real-valued functions of the variable 
$\xi =t-vx$. Here, $v$ and $k$ are real constants. Indeed, substituting Eq. (\ref{ansatz}) into Eq. (\ref{CCQGLE}) and separating the real and imaginary parts of the so-obtained equation, one arrives at the following system of nonlinear ordinary differential equations 
\begin{equation} \label{Eq1}
	\rho ^{\prime \prime }-4 \rho ^{\prime }\left( \frac{\alpha \tau _{R}\rho ^{2}-\beta v}{\alpha ^{2}+4\beta ^{2}}\right)
	+\frac{\rho
		\phi ^{\prime }\left[ 2\alpha v-\left( \alpha ^{2}+4\beta ^{2}\right) \phi
		^{\prime }\right] }{\alpha ^{2}+4\beta ^{2}}
	+ \frac{2(2\beta \delta +\alpha k)\rho }{\alpha ^{2}+4\beta ^{2}}+\frac{%
		2(\alpha \gamma +2\beta \varepsilon )\rho ^{3}}{\alpha ^{2}+4\beta ^{2}}+\frac{%
		2(\alpha \lambda +2\beta \mu )\rho ^{5}}{\alpha ^{2}+4\beta ^{2}}=0,
\end{equation}
\begin{equation} \label{Eq2}
	\phi ^{\prime \prime }
	+\frac{4\beta v\phi ^{\prime }}{\alpha ^{2}+4\beta ^{2}	}
	-\rho ^{\prime }\left[ \frac{8\beta \tau _{R}\rho }{\alpha ^{2}+4\beta ^{2}}
	+\frac{2\alpha v-2\left( \alpha ^{2}+4\beta ^{2}\right) \phi ^{\prime }}{\left( \alpha ^{2}
		+4\beta ^{2}\right) \rho }\right]
	-\frac{2\alpha \delta-4\beta k}{\alpha ^{2}+4\beta ^{2}}
	-\frac{(2\alpha \varepsilon -4\beta \gamma)\rho ^{2}}{\alpha ^{2}+4\beta^{2}}
	-\frac{(2\alpha \mu -4\beta \lambda )\rho
		^{4}}{\alpha ^{2}+4\beta ^{2}}=0, 
\end{equation}
\noindent for the
functions $\rho \left( \xi \right)$ and $\phi \left( \xi \right)$, where the primes denote differentiation with respect to the variable $\xi$. Let us remark that for $\alpha=1$ this system is equivalent (up to some differences in the notation) to the relevant systems of equations obtained in \cite{Nisha2020} and \cite{Uzunov2014}.

\section{Differential Constraints}
\label{Sec:3}

Looking for solutions to the system of equations (\ref{Eq1}) and (\ref{Eq2}), we enlarge it by adding the additional differential equation 
\begin{equation} \label{DC}
	\phi^{\prime }=\frac{2\tau _{R}\beta }{\alpha ^{2}+4\beta ^{2}} \rho^{2}
	+\frac{\alpha \left( \alpha \varepsilon -2\beta \gamma \right) }{4\tau _{R}\beta ^{2}},
\end{equation}
and its differential consequence
\begin{equation} \label{DCons}
	\phi ^{\prime \prime }=\frac{4\tau _{R}\beta }{\alpha ^{2}+4\beta ^{2}}\rho
	\rho ^{\prime }
\end{equation}
as differential constraints. Then, it is straightforward to verify that after eliminating $\phi ^{\prime }$ and $\phi ^{\prime \prime }$
from Eqs. (\ref{Eq1}) and (\ref{Eq2}) using the above  expressions (\ref{DC}) and (\ref{DCons}), respectively, and assuming
\begin{equation}
	v=\frac{\left( \alpha ^{2}+4\beta ^{2}\right) \left( \alpha \varepsilon -2\beta \gamma \right) }{4\tau _{R}\beta ^{2}}, \quad
	k=\frac{\alpha \delta }{2\beta }-\frac{\alpha \left( \alpha	^{2}+4\beta ^{2}\right) \left( \alpha \varepsilon -2\beta \gamma \right) ^{2}}{16\tau _{R}^{2}\beta ^{4}},\quad
	\mu =\frac{2\beta \lambda }{\alpha },
	\label{Coeff1}
\end{equation}
the regarded system reduces to the single equation for the function $\rho(\xi)$ only, viz.
\begin{equation} \label{Eq}
	\rho ^{\prime \prime }+\left( \frac{\alpha \varepsilon -2\beta \gamma }{\tau
		_{R}\beta }-\frac{4\alpha \tau _{R}}{\alpha ^{2}+4\beta ^{2}}\rho
	^{2}\right) \rho ^{\prime }+\left[ \frac{\delta }{\beta }-\frac{\alpha
		^{2}\left( \alpha \varepsilon -2\beta \gamma \right) ^{2}}{16\tau
		_{R}^{2}\beta ^{4}}\right] \rho +\frac{2\alpha \gamma +4\beta \varepsilon }{%
		\alpha ^{2}+4\beta ^{2}}\rho ^{3}+\left( \frac{2 \lambda}{\alpha }-\frac{4\tau
		_{R}^{2}\beta ^{2}}{\left( \alpha ^{2}+4\beta ^{2}\right) ^{2}}\right) \rho
	^{5}=0.
\end{equation}
Thus, each couple of functions $\rho(\xi)$ and $\phi (\xi )$ such that $\rho(\xi)$ is a solution of equation (\ref{Eq}) and
\begin{equation} \label{Solphi}
	\phi (\xi )=\frac{2\tau _{R}\beta }{\alpha ^{2}+4\beta ^{2}}\int \rho(\xi)
	^{2}d\xi +\frac{\alpha (\alpha \varepsilon -2\beta \gamma )}{4\tau
		_{R}\beta ^{2}}\xi +\phi _{0},
\end{equation}
where $\phi _{0}$ is a real constant, is a solution of the system of equations (\ref{Eq1}) and (\ref{Eq2}) provided that the conditions (\ref{Coeff1}) hold. Let us remark that the expression (\ref{Solphi}) of the function $\phi (\xi)$ is obtained by integrating Eq. (\ref{DC}).

Now, in order to find some particular solutions to equation (\ref{Eq}) we look for the conditions under which this equation is compatible with a sub-equation of the form
\begin{equation} \label{LWDC}
	\rho^{\prime }=c \left( a-\rho \right)^2 \left(b-\rho \right),
\end{equation}
where $a$, $b$ and $c$ are real constants. Notice that Eq. (\ref{LWDC}) is nothing but a differential constraint imposed on Eq. (\ref{Eq}). The choice of this equation as the sub-equation is motivated by the fact that its real-valued solutions can be given in closed form in terms of the principle branch $W_0$ of the Lambert $W$ function \cite{Corless1996} (see the Appendix for details).

Substitution of  Eq. (\ref{LWDC})
and its differential consequence
\begin{equation}\label{LWDC1}
	\rho ^{\prime \prime }=-3c^{2}\left( a-\rho \right) ^{3}\left( b-\rho
	\right) \left( \frac{a+2b}{3}-\rho \right) 
\end{equation}
into Eq. (\ref{Eq}) leads to a quintic polynomial in the function $\rho(\xi)$ that should be equated to zero. Equating to zero the coefficients of this polynomial, one gets to a system of six algebraic equations for the parameters  $a$, $b$, $c$, $\tau _{R}$, $\alpha$, $\beta$, $\gamma$, $\delta$, $ \varepsilon $ and $\lambda$. Notice that the rest of the parameters, namely $v$, $k$ and $\mu$, are already specified via relations (\ref{Coeff1}). If these parameters can be chosen so that the aforementioned system of algebraic equations is satisfied, then 
any solution of the respective equation of the form (\ref{LWDC}) is a solution of the corresponding equation of the form (\ref{Eq}) as well. Following the procedure described above, four such cases are identified below. 
All the results presented in the next Section can be easily verified by direct computation bearing in mind Eqs. (\ref{LWF}), (\ref{DLWF}), (\ref{GenLWKSol}) and (\ref{Int}).

\section{Exact Solutions}
\label{Sec:4}

\begin{proposition}
Let
\begin{equation}\label{Case1}
	b=-a \quad \mathrm{and} \quad c=-\frac{4\tau _{R}\alpha }{5\left( \alpha ^{2}+4\beta^{2}\right) }. 
\end{equation}
Then, under the following relations between the parameters, namely 
\begin{equation}\label{PConds1}
	\lambda =\frac{2\tau _{R}^{2}\alpha \left( 8\alpha^{2}+25\beta ^{2}\right) }{25\left( \alpha ^{2}+4\beta ^{2}\right) ^{2}},\quad
	\varepsilon =-\frac{44a^{2}\tau _{R}^{2}\alpha ^{2}\beta }{25\left(\alpha ^{2}+4\beta ^{2}\right) ^{2}},\quad
	\gamma =-\frac{8a^{2}\tau_{R}^{2}\alpha \left( 4\alpha ^{2}+5\beta ^{2}\right) }{25\left( \alpha^{2}+4\beta ^{2}\right) ^{2}},\quad
	\delta =\frac{a^{4}\tau _{R}^{2}\alpha	^{2}\left( \alpha ^{2}+32\beta ^{2}\right) }{25\beta \left( \alpha^{2}+4\beta ^{2}\right) ^{2}},
\end{equation}
Eqs. (\ref{Eq}) and (\ref{LWDC}) are compatible and have common solutions of the form
\begin{equation} \label{rhosol1}
	\rho \left( \xi \right) =a\left( 1-\frac{2}{1+W_0 \left( \exp [-4ca^{2}\xi -\xi _{0}]\right)} \right),
\end{equation}
where $\xi _{0}$ is a real constant, cf. (\ref{GenLWKSol}).
Consequently, according to the expressions (\ref{Solphi}) and (\ref{Int})
\begin{equation}  \label{phisol1}
	\phi (\xi )=-\frac{5 \beta }{8 \alpha }
	\left( \ln \left[ \frac{\left( 1+Q\left( \xi \right) \right) ^{4}}{Q\left(
		\xi \right) }\right] -Q\left( \xi \right) \right)
	+\frac{a^2 \tau _{R} \alpha^2}{5 \beta (\alpha^2+ 4\beta ^{2})}
	\xi +\phi _{0}, \quad
		Q\left( \xi \right) =W_0 \left( \exp [-4ca^{2}\xi -\xi _{0}]\right).
\end{equation}
\end{proposition}
It should be remarked that the sub-case in which $\alpha =1$ and $c=-1$ is regarded and analyzed in the context of nonlinear optics in the recent work by Nisha \textit{et al.} \cite{Nisha2020}.
In this sub-case, however, there is one more relation between the parameters, namely
$	4\tau _{R}=5\left( 1+4\beta^{2}\right) .$
    	
\begin{proposition}
Let
\begin{equation}\label{Case2}
	b=-2a, \quad c=-\frac{4\tau _{R}\alpha }{3\left( \alpha ^{2}+4\beta^{2}\right) }. 
\end{equation}
Then, Eqs. (\ref{Eq}) and (\ref{LWDC}) are compatible under the conditions
\begin{equation}\label{PConds2}
	\lambda =\frac{2\tau _{R}^{2}\alpha \beta ^{2}}{\left(
		\alpha ^{2}+4\beta ^{2}\right) ^{2}},\quad \varepsilon =\frac{4a^{2}\tau
		_{R}^{2}\alpha ^{2}\beta ^{2}}{\left( \alpha ^{2}+4\beta ^{2}\right) ^{2}}%
	,\quad \gamma =-\frac{8a^{2}\tau _{R}^{2}\alpha \beta ^{2}}{\left( \alpha
		^{2}+4\beta ^{2}\right) ^{2}},\quad \delta =\frac{a^{4}\tau _{R}^{2}\alpha
		^{4}}{\beta \left( \alpha ^{2}+4\beta ^{2}\right) ^{2}},
\end{equation}
and have common solutions of the form 
\begin{equation} \label{rhosol2}
	\rho \left( \xi \right) =a\left( 1-\frac{3}{1+W_0 \left( \exp [-9ca^{2}\xi -\xi	_{0}]\right) }\right),
\end{equation}
where $\xi _{0}$ is a real constant, see (\ref{GenLWKSol}).
Consequently, according to the expressions (\ref{Solphi}) and (\ref{Int}),
\begin{equation}  \label{phisol2}
	\phi (\xi )=
	-\frac{\beta }{6 \alpha }%
	\left( \ln \left[ \frac{\left( 1+Q\left( \xi \right) \right) ^{9}}{Q\left(
		\xi \right)^{4} }\right] -Q\left( \xi \right) \right)+\frac{a^2 \tau _{R} \alpha^2}
	{\beta (\alpha^2+ 4\beta ^{2})}
	\xi +\phi _{0}, \quad
	Q\left( \xi \right) =W_0 \left( \exp [-9ca^{2}\xi -\xi_{0}]\right).
\end{equation}
\end{proposition}
\begin{proposition}
Let
\begin{equation}\label{Case3}
	b=0,\quad c=-\frac{4\tau _{R}\alpha }{5\left( \alpha ^{2}+4\beta ^{2}\right)}.
\end{equation}
Then, Eqs. (\ref{Eq}) and (\ref{LWDC}) are compatible under the conditions
\begin{equation}\label{PConds3}
	\lambda =\frac{2\tau _{R}^{2}\alpha \left( 8\alpha ^{2}+25\beta
		^{2}\right) }{25\left( \alpha ^{2}+4\beta ^{2}\right) ^{2}},\quad
	\varepsilon =-\frac{124a^{2}\tau _{R}^{2}a^{2}\beta }{25\left( \alpha
		^{2}+4\beta ^{2}\right) ^{2}},\quad \gamma =\frac{8a^{2}\tau _{R}^{2}\alpha
		\left( 15\beta ^{2}-4\alpha ^{2}\right) }{25\left( \alpha ^{2}+4\beta
		^{2}\right) ^{2}},\quad \delta =\frac{a^{4}\tau _{R}^{2}\alpha ^{2}\left(
		9\alpha ^{2}+32\beta ^{2}\right) }{25\beta \left( \alpha ^{2}+4\beta
		^{2}\right) ^{2}}, 
\end{equation}
and have common solutions of the form
\begin{equation} \label{rhosol3}
	\rho \left( \xi \right) =a\left( 1-\frac{1}{1+W_0 \left( \exp [-ca^{2}\xi -\xi _{0}]\right)} \right),
\end{equation}
where $\xi _{0}$ is a real constant, see (\ref{GenLWKSol}). Consequently, according to the expressions (\ref{Solphi}) and (\ref{Int}),
\begin{equation}  \label{phisol3}
	\phi (\xi )=\frac{5 \beta }{2 \alpha }
	\left(Q\left( \xi \right)- \ln \left[ 1+Q\left( \xi \right) \right] \right)
	-\frac{3a^2 \tau _{R} \alpha^2}{5 \beta (\alpha^2+ 4\beta ^{2})}
	\xi +\phi _{0}, \quad
		Q\left( \xi \right) =W_0 \left( \exp [-ca^{2}\xi -\xi _{0}]\right). 
\end{equation}
\end{proposition}
\begin{proposition}
Let
\begin{equation}\label{Case4}
	a=0,\quad c=-\frac{4\tau _{R}\alpha }{5\left( \alpha ^{2}+4\beta ^{2}\right)}.
\end{equation}
Then, under the conditions
\begin{equation} \label{PConds4}
	\lambda =\frac{2\tau _{R}^{2}\alpha \left( 8\alpha ^{2}+25\beta
		^{2}\right) }{25\left( \alpha ^{2}+4\beta ^{2}\right) ^{2}},\quad
	\varepsilon =-\frac{32b^{2}\tau _{R}^{2}\alpha ^{2}\beta }{25\left( \alpha
		^{2}+4\beta ^{2}\right) ^{2}},\quad \gamma =-\frac{16b^{2}\tau
		_{R}^{2}\alpha ^{3}}{25\left( \alpha ^{2}+4\beta ^{2}\right) ^{2}},\quad
	\delta =0,
\end{equation}
Eqs. (\ref{Eq}) and (\ref{LWDC}) are compatible and have common solutions of the form
\begin{equation} \label{rhosol4}
	\rho \left( \xi \right) = \frac{b}{1+W_0 \left( \exp [-cb^{2}\xi -\xi	_{0}] \right) },
\end{equation}
where $\xi _{0}$ is a real constant, see Eq. (\ref{GenLWKSol}).
In this case, relations (\ref{Solphi}) and (\ref{Int}) imply
\begin{equation} \label{phisol4}
	\phi (\xi )=\frac{5\beta }{2\alpha }\ln \left[ \frac{Q\left( \xi \right) }{	1+Q\left( \xi \right) }\right] +\phi _{0},  \quad
	Q\left( \xi \right) =W_0 \left( \exp [-cb^{2}\xi -\xi	_{0}] \right).
	\end{equation}
\end{proposition}
\begin{figure}[h]
	\centering
	\includegraphics[width=5.35cm]{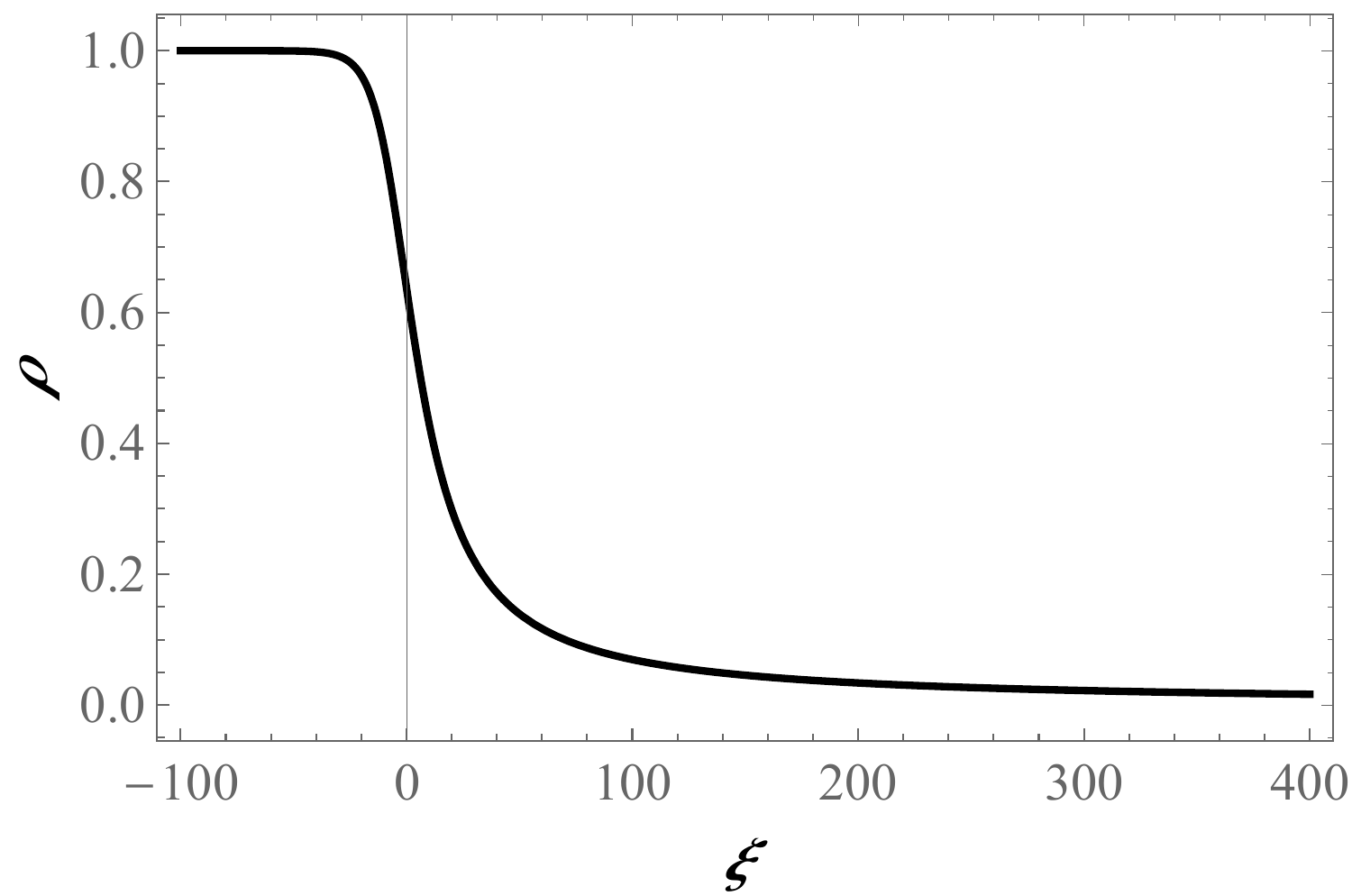}
	\includegraphics[width=5.4cm]{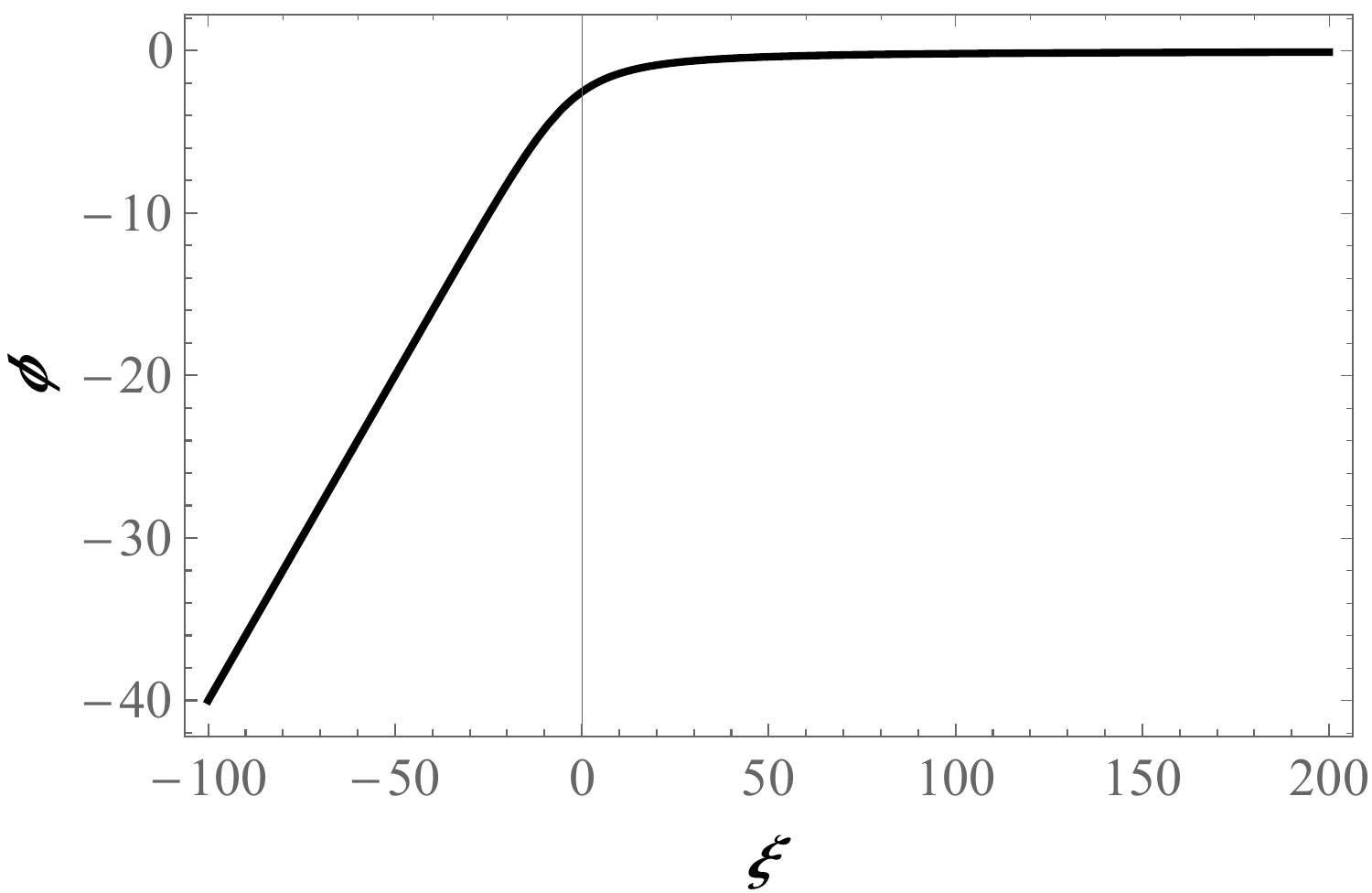}
	\includegraphics[width=5.4cm]{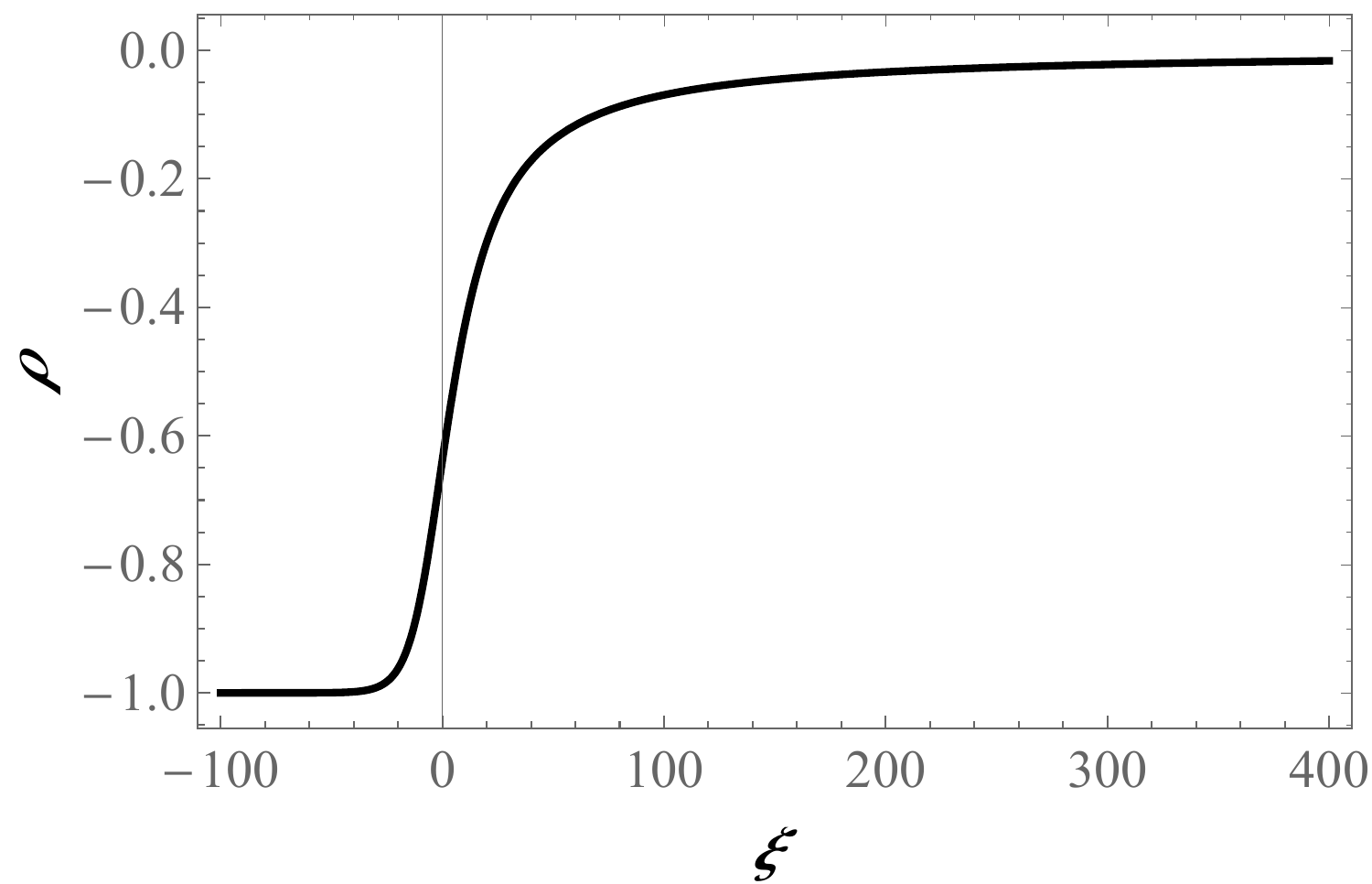}
	\caption{Kink-like solutions to  Eq. (\ref{Eq}) of the form (\ref{rhosol4}): (left) $b=1$ ; (right) $b=-1$; (middle) graph of the respective function $\phi (\xi )$ of the form (\ref{phisol4}) for $b=\pm 1$. Here, $\xi_0=\phi_0 = 0$, $\alpha=\beta=\tau_{R}=1$, the values of the rest of the parameters being determined via formulas (\ref{Coeff1}), (\ref{Case4}) and (\ref{PConds4}).
	}
	\label{Fig1}
\end{figure}

\section{Concluding Remarks}

The paper reports results of a study of the family of complex cubic-quintic Ginzburg--Landau equations (\ref{CCQGLE}) for which $\tau _{R} \neq 0$, $\alpha \neq 0$ and $\beta \neq 0$. 
These equations admit reduction to the system of ordinary differential equations (\ref{Eq1}) and (\ref{Eq2}) through the ansatz (\ref{ansatz}).
This system of ordinary differential equations is solved by imposing the differential constraints (\ref{DC}) and (\ref{LWDC}).
Four classes of equations of the form (\ref{CCQGLE}) that are compatible with the foregoing differential constraints are presented in Section 4, see Propositions 1, 2, 3 and 4. In each such case, the solution $U(x,t)$ of the respective equation (\ref{CCQGLE}) is expressed in explicit analytic form  by means of the Lambert $W$ function via the functions $\rho(\xi)$, $\varphi(\xi)$ and relation (\ref{ansatz}). Typical examples of kink-like solutions are depicted in Figure \ref{Fig1}.

\section*{Acknowledgements}
	
This work has been accomplished with the financial support by Grant No BG05M2OP001-1.002-0011-C02 financed by the Science and Education for Smart Growth Operational Program (2014--2020) and co-financed by the European Union through the European structural and investment funds.
The author would like to acknowledge also the support from the Bulgarian Science Fund under grant K$\Pi$-06-H22/2.

\section*{Appendix. Solution of equation (\ref{LWDC}) in terms of the Lambert W function}

Let us recall that the Lambert $W$ function \cite{Corless1996} is defined to be the function that solves the transcendental equation
\renewcommand\theequation{{A1}}  
\begin{equation} \label{LWF}
	W(z) \exp[W(z)]=z.
\end{equation}
This is a multivalued function, but here we are interested in bounded real-valued solutions of Eq. (\ref{LWF}) when its argument $z$ is a positive real number. In this case, Eq. (\ref{LWF}) have only one real-valued solution given by the principal branch $W_0$ of the Lambert $W$ function and $W_0 \in [0, \infty )$. Let us also recall that the derivative of the function $W$ is given by the expression
\renewcommand\theequation{{A2}}
\begin{equation}  \label{DLWF}
	\frac{dW(z)}{dz}=\frac{  \exp[-W(z)]}{1 +W(z) }.
\end{equation}
Now, we are ready to express in closed form the real-valued solutions of Eq. (\ref{LWDC}), that is
\[
	\rho^{\prime }=c \left( a-\rho \right)^2 \left(b-\rho \right).
\]

Under the change of the dependent variable $\rho$ of the form 
\renewcommand\theequation{{A3}}
\begin{equation}\label{TrEq}
	\rho=\frac{a w+b}{w +1},
\end{equation}
Eq. (\ref{LWDC}) transforms into the equation
\renewcommand\theequation{{A4}}
\begin{equation} \label{TrLWEq}
	w^{\prime }=-\frac{c (a-b)^2 w}{w+1}
\end{equation}
whose general solution can easily be expressed implicitly in the form
\renewcommand\theequation{{A5}}
\begin{equation} \label{TrLWEq_ISol}
	|w| \exp[ w]= \exp[-c (a-b)^2 \xi-\xi_0],
\end{equation}
where $\xi_0$ is a constant of integration. Thus, according to the definition (\ref{LWF}), the general bounded real-valued solution of Eq. (\ref{TrLWEq}) reads
\renewcommand\theequation{{A6}}
\begin{equation} \label{TrLWEq_ISolW}
	w(\xi) = W_0 \left( \exp[-c (a-b)^2 \xi-\xi_0]\right)
\end{equation}
since $ \exp[-c (a-b)^2 \xi-\xi_0]>0$ for each $\xi \in (-\infty,\infty)$.
Finally, substituting the expression (\ref{TrLWEq_ISolW}) into Eq. (\ref{TrEq}) one obtains the sought solution of Eq. (\ref{LWDC}), viz.
\renewcommand\theequation{{A7}}
\begin{equation} \label{GenLWKSol}
	\rho \left( \xi \right) = a+\frac{b-a}{1+W_0 \left( \exp [-c(a-b)^{2}\xi -\xi _{0}]\right)}.
\end{equation}
Then, it is easy to verify using formula (\ref{DLWF}) that
\renewcommand\theequation{{A8}}  
\begin{equation}\label{Int}
	\int \rho(\xi) ^{2}d\xi =-\frac{1}{c(a-b)^{2}}\left( a^{2}+\ln \left[ \frac{Q\left( \xi \right)
		^{b^{2}}}{\left( 1+Q\left( \xi \right) \right) ^{(a-b)^{2}}}\right] \right)
	+ const, \quad
		Q(\xi) = W_0 \left( \exp[-c (a-b)^2 \xi-\xi_0]\right).
\end{equation}

\end{document}